\newcommand{\tlp}[1]{#1} %
\newcommand{\nottlp}[1]{}  %

\newcommand{\aaai}[1]{} %

\newcommand{\arxiv}[1]{} %

\newcommand{\full}[1]{} %
\newcommand{\short}[1]{} %

\newcommand{\oldtlp}[1]{} %
\arxiv{}
\aaai{}
\tlp{
  \newcommand{\sepexpe}[1]{#1}
  \newcommand{\notsep}[1]{}
}

\arxiv{}

\tlp{
\documentclass{tlp}
\renewcommand{\cite}[1]{\citep{#1}}
\usepackage{graphicx} %
\usepackage{hyperref} %
}

\aaai{} %

\usepackage{alltt}
\usepackage{xspace}

\nottlp{}

\usepackage{caption}
\usepackage{subcaption}

\newcommand{\Hex}[1]{\hspace{#1ex}}
\newcommand{\Vex}[1]{\vspace{#1ex}}

\newcommand{\mysec}[1]{\Vex{\tlp{-0}\nottlp{}}\section{#1}\Vex{\tlp{-0}\nottlp{}}}
\newcommand{\mysubsec}[1]{\Vex{\tlp{-0}\nottlp{}}\subsection{#1}\Vex{\tlp{-0}\nottlp{}}}
\newcommand{\mypar}[1]{\Vex{\tlp{-0}\nottlp{}}\paragraph{\bf #1.}}
\newcommand{\myendneg}{\Vex{\tlp{-1.5}\nottlp{}}}

\newenvironment{code}{\Vex{\tlp{0}\nottlp{}}\begin{alltt}\footnotesize}{\end{alltt}\Vex{\tlp{-.5}\nottlp{}}}
\NewDocumentCommand{\co}{+m}{\mbox{\footnotesize\tt #1}} %
\newcommand{\m}[1]{$#1$} %
\newcommand{\gname}[2]{\m{\co{#1}_{\co{#2}}}}
\newcommand{\souf}{Souffl\'e\xspace}

\arxiv{}

\aaai{}%

\begin{document}

\tlp{
\lefttitle{Y.A.\ Liu, J.\ Idogun, S.D.\ Stoller, and Y.\ Tong}
\jnlPage{1}{17} %
\jnlDoiYr{2026}
\doival{10.1017/xxxxx}

\oldtlp{} %

\title[]{
  Efficiency of Analysis of Transitive Relations
  using Query-Driven, Ground-and-Solve, and Fact-Driven Inference\Vex{-0}}

\begin{authgrp}
\author{Yanhong A.\ Liu \Hex{1.5} John Idogun \Hex{1.5} 
  Scott D.\ Stoller \Hex{1.5} 
  Yi Tong}
\affiliation{Stony Brook University\\
  {\tt \{liu,jidogun,stoller,yittong\}@cs.stonybrook.edu}\Vex{-2}\\\Vex{-0}}
\end{authgrp}
}%

\maketitle

\begin{abstract}
  Logic rules allow analysis of complex relationships to be expressed
  easily, especially for transitive relations in critical applications.
  However, understanding and predicting the efficiency of different
  inference methods remain challenging, even for simplest rules given
  different kinds of input data.

  This paper analyzes the efficiency of all three types of well-known
  inference methods---query-driven, ground-and-solve, and
  fact-driven---along with their respective optimizations, and compares
  with optimal complexities for the first time, for analyzing transitive
  graph relations.  We also experiment with rule systems widely considered
  to have the best performance.  We analyze all well-known rule variants
  and widely varying input graphs.  The results include precisely
  calculated optimal time complexities; comparative analysis across
  different inference methods, rule variants, and graph types; confirmation
  with performance experiments; as well as discovery of a performance bug.

\end{abstract}
\myendneg

\mysec{Introduction}

With rapid advances of LLMs in AI, but without assurance in their speedy
question answering, rigorous logic inference has become even more important
for critical applications.  However, understanding drastically different
inference methods and their efficiency on different kinds of input data has
remained a significant challenge.

Logic rules allow analysis of complex relationships to be expressed easily
and precisely, especially with recursive rules for transitive relations in
critical applications, e.g., understanding
program structures, flows, and
dependencies~\cite{smara15pointer,flores2020datalog};
enforcing
security policies~\cite{%
    LiMit03,Hri+07SPKI-PPDP}; probing 
networking systems including social
networks~\cite{loo09decl,seo2013socialite}; and analyzing
knowledge graphs in
general~\cite{bellomarini2018vadalog,hogan2021knowledge}.
\aaai{}

Powerful rule systems allow analysis 
expressed using rules to be executed automatically without
low-level programming.  There are not only Prolog systems since the
1970s~\cite{Sterling:Shapiro:94} and answer-set programming (ASP) systems
since the 1990s~\cite{GebKKS12book}, but also more restricted Datalog
systems since the 1980s~\cite{maier18hist-wbook} that have been regularly
(re)created since the 2000s~\cite{ketsman2022modern}
due to increasing need of 
analysis,
especially for transitive relations.

How to best utilize the power and efficiency of automatic inference in rule
systems?  It is regularly observed that different inference methods and
rule systems can have significantly different performance, besides
requiring different ways to express the analysis:
\begin{itemize}

\item %
  The same analysis %
  written differently using rules,
  as well as the same size of input data having different shapes, so to
  speak, can have drastically
  different running times.  Different inference methods can tolerate these
  differences very differently.

\item For applications written using rule languages together with other
  languages, where input data and analysis results are large, efficiency of
  passing data and results can also make a large difference, and may
  dominate the running time.

\end{itemize}
It is desirable to better understand and predict the efficiency of
different inference methods, for different variants of rules running on
different kinds of input data of different sizes.

This paper presents a systematic, precise, and detailed comparative
analysis of the efficiency of computing transitive relations.  We consider
all three types of well-known inference methods---query-driven,
ground-and-solve, and fact-driven---along with their respective
optimizations, across different recursive rules and different input graph
types, and compare with optimal complexities for the first time.

We also confirm analyzed results with experiments using rule systems widely
considered to have best performances: Prolog system
XSB~\cite{SagSW94xsb,xsb22} for query-driven, ASP system
clingo~\cite{gebser2019multi,clingo26} for ground-and-solve, and Datalog
system \souf~\cite{jordan2016souffle,souffle26github} for fact-driven.

We use the most well-known rules for computing the transitive closure of
arbitrary graphs, but our method of calculating optimal complexities and
our analyses for all three inference methods and rule variants are general.
\begin{itemize}
\item The rules for transitive closure are also known for solving graph
  reachability, the most fundamental problem in analyzing
  deep relationships, especially with unbounded depth, which
  makes transitive closure well known to be beyond first-order
  logic~\cite{gradel1991transitive}.

\item The rules capture both key features of analyses of transitive
  relations---join and recursion. Rules with more joins and recursions,
  over more relations with more arguments, reduce to join and recursion as
  in the rules for transitive closure~\cite{%
    LiuSto09Rules-TOPLAS,Liu13book}.

\item The rules have all three variants of recursion---left, right, and
  double---and their significant impact on efficiency has not been analyzed
  in prior comparisons
  except for showing two curves on one ad hoc graph
  type~\cite{Liu+22RuleLang-arxiv,Liu+23RuleLangBench-ICLP}.
\end{itemize}
We also examine a wide variety of input graph types, including all from an
extensive previous study~\cite{brass2019performance}.

We present precisely calculated optimal time complexities for all
combinations of rule variants and graph types; comparative analysis across
different inference methods for all combinations; and confirmation with
detailed
experiments.
In particular, we show how the analyzed complexities explain the sameness
and differences in actual query times precisely.
These results help predict the efficiency of different inference methods
and best utilize different rule systems.  We also discovered a performance
bug in XSB.

There has been significant bodies of work on both precise complexities and
performance measurements, as discussed in Section~\ref{sec-related}.  Our
contributions include
four main aspects that are studied systematically for the first time.
\begin{itemize}

\item A comparative analysis across all three types of well-known inference
  methods, especially with their respective optimizations, and their impact
  on the complexity and efficiency of analysis of transitive relations.

\item Precise optimal time complexity calculation for inference for %
  different rule variants and input graph types, not ignoring rule variants
  and considering only
  data sizes.

\item Detailed running time measurements for all of compiling and loading
  programs, reading input data, querying or grounding and solving, and
  writing query results.

\item Confirmation of measured times against analyzed efficiency across all
  inference methods, rule variants, and graph types, helping best utilize
  different rule systems and possible improvements.

\end{itemize}
\short{}

The rest of the paper is organized as follows. Section~\ref{sec-sys}
compares different inference methods for logic rules and different rule
systems and languages supporting the different inference methods.
Section~\ref{sec-rules-graphs} explains the rule variants and graph types
analyzed and evaluated.
Section~\ref{sec-analyze} presents precise optimal time complexity analysis
as well as analysis for different inference and optimization methods.
Section~\ref{sec-expe} describes experimental results with key findings
and analysis.
Section~\ref{sec-related} discusses related work, 
and as conclusion, directions for future work.
\arxiv{}

\mysec{Comparing inference methods and rule systems}
\label{sec-sys}

We consider all three types of drastically different inference methods,
used by
mostly separate communities of logic rule systems:
\begin{enumerate}

\item query-driven inference, also known as goal-directed search or
  top-down evaluation, the core of Prolog systems and
  variants~\cite{Sterling:Shapiro:94};

\item ground-and-solve inference, the core of ASP %
  systems~\cite{GebKKS12book};

\item fact-driven inference, also called bottom-up evaluation, the core of
  Datalog systems and variants~\cite{maier18hist-wbook}.
\end{enumerate}
We distill and contrast the fundamentals of these different methods, along
with their
powerful optimizations to overcome their root sources of inefficiency.
Their impacts on the efficiency of analysis of transitive relations are
described in Sections~\ref{sec-analyze} and~\ref{sec-expe}.

\mysubsec{Inference methods with powerful optimizations}

Despite significant differences, all inference methods are aimed at
providing the best efficiency.
It is the optimized inference that is to be compared with optimal
complexities.

\mypar{Query-driven inference} 
Query-driven inference performs top-down evaluation starting from the given
query, searching through rules recursively---querying, in left-to-right
order, hypotheses of rules whose conclusion matches the current query%
---until supporting facts for all queries are found.
This goal-directed search allows rules to be used for not only logic
programming, but also relational programming as with database languages and
recursive programming as with functional
languages~\cite{Sterling:Shapiro:94}.

However, simple top-down evaluation may not terminate for recursive rules
over cyclic relationships, and may take exponential time on acyclic
relationships. This is because the same queries may be repeated infinitely
or exponentially many times. Top-down evaluation with tabling solves this
fundamental problem by remembering the queries performed and reusing the
query results~\cite{TamSat86,CheWar96,swift2012xsb}.

\mypar{Ground-and-solve inference}
Ground-and-solve inference finds answers to queries
by using grounding followed by solving~\cite{kaufmann2016grounding}.
Grounding instantiates rules, replacing variables in rules with values that
contain no variables, yielding a propositional program.  Solving then finds
satisfying answers to the propositional program under 
stable model semantics, 
by using try-and-backtrack combinatorial search.

However, naive grounding and solving can lead to combinatorial explosion
and 
even nontermination.
Optimized grounding tries to consider only possible combinations of values
that can match given or derived facts, not all
combinations~\cite{kaufmann2016grounding}.  Advanced search uses
conflict-driven clause learning (CDCL) with back-jumping instead of
backtracking, skipping more areas of unfruitful
search~\cite{marques1999grasp,GebKKS12book}.

\mypar{Fact-driven inference}
Fact-driven inference performs bottom-up evaluation starting from facts,
repeatedly applying rules---inferring new facts from conclusions of rules
whose hypotheses match existing facts---until no more new facts can be
inferred.
This is a least fixed-point computation, with a larger set of facts
inferred in each iteration using facts from the previous iteration, until a
fixed point is reached.

Naive bottom-up inference from all facts in each iteration can result in
heavily repeated computation.  Well-known semi-naive evaluation considers
only new facts added in each iteration, yielding drastic efficiency
improvement~\cite{bancilhon1986naive}.  However, it may still have
redundant computations in each iteration.  Optimal computation using
incrementalization with minimum increment considers only one fact in each
iteration and avoids all redundant computations, and furthermore gives
precise complexity
guarantees~\cite{\full{}LiuSto09Rules-TOPLAS,Liu13book}.

\mysubsec{%
Rule systems and languages}
For confirming analyzed inference efficiency through experiments, we use
state-of-the-art rule systems
widely considered to have the best performance using different inference
methods:
Prolog system XSB~\cite{SagSW94xsb,xsb22} %
for query-driven,
ASP system clingo\cite{gebser2019multi,clingo26} for ground-and-solve, and
Datalog system \souf~\cite{jordan2016souffle,souffle26github} for
fact-driven.

\begin{table}[t]
  \tlp{\footnotesize}\nottlp{}
  \centering
\tlp{\renewcommand{\arraystretch}{0}}
\begin{tabular}{@{\arxiv{}}p{\tlp{19}\aaai{}\arxiv{}ex}@{~}||@{\arxiv{}}p{\tlp{25}\aaai{}\arxiv{}ex}@{~}|@{\arxiv{}}p{\tlp{33}\aaai{}\arxiv{}ex}@{~}|@{\arxiv{}}p{\tlp{24}\aaai{}\arxiv{}ex}@{\arxiv{}}}%
               & Query-driven & Ground-and-solve & Fact-driven\\\hline\hline
  inference method & top-down evaluation
                              & propositions\arxiv{} and satisfiability
                                                 & bottom-up evaluation\\\hline

  optimized\newline inference & tabling 
                              & optimized grounding and\tlp{\newline} CDCL
                                           & semi-naive;\newline
                                             minimum\aaai{} increment\\
  \hline
  rule systems & Prolog and\aaai{} variants 
                              & ASP %
                                           & Datalog and\aaai{} 
                                             variants\\\hline
  viewed as\newline most efficient
               & ~\newline XSB  & ~\newline clingo   & ~\newline \souf\\\hline
  unique\newline advantages   & general programming,\newline
                              query-\aaai{}restricted search
                              & constraint programming and \tlp{\newline} solving %
                                           & generating standalone %
                                             code from rules\\\hline
\end{tabular}
\caption{Inference methods and rule systems.\Vex{-0}}
\label{tab-sys}
\end{table}
\mypar{Unique advantages}
Beyond their shared power,
rule systems implementing different inference methods have different unique
advantages that are critical for different kinds of applications.
\begin{itemize}

\item Prolog systems support general-purpose programming. Also, their
  goal-directed search can answer queries (e.g., only paths from a
  particular vertex) much more efficiently by inferring only needed facts,
  not all facts.

\item ASP systems allow powerful constraint programming and solving for
  constraint satisfaction and optimization problems (e.g., the traveling
  salesman problem).

\item Datalog systems often
  generate standalone code specialized for the given rules (e.g., 
  code for computing transitive closure) without requiring a general
  inference system at runtime, e.g., \souf does so, but XSB and clingo do
  not.

\end{itemize}
Table~\ref{tab-sys} summarizes the inference methods and rule systems.
Note that 
other systems, 
e.g., Ciao and GNU Prolog, can also generate standalone code and support
CLP for writing constraints, but they are less efficient;
GNU Prolog even requires manual programming for tabling to avoid
nontermination or exponential time for recursive rules on graphs.

\mypar{Rule languages} %
Largely due to different inference methods used, rule languages supporting
these methods have different features.  We show this using the 
transitive closure problem.

Given a predicate, \co{edge}, that asserts whether there is an edge from a
first vertex to a second vertex, the transitive closure problem defines a
predicate, \co{path},
that asserts whether there is a path 
from a first vertex 
to a second vertex by following the edges.
This can be expressed in all dominant logic languages, including Prolog,
ASP, and Datalog variants, as follows:
\begin{code}
  path(X,Y) :- edge(X,Y).
  path(X,Y) :- edge(X,Z), path(Z,Y).
\end{code}
Often, the name \co{reachable} is used in place of \co{path}.

Besides the two rules above, different rules languages require different
additional specifications.
\begin{itemize}

\item 
  {\bf Prolog systems like XSB.} To avoid nontermination or exponential
  time, tabling directives are needed, as follows for using default tabling
  for predicate \co{path}:
\begin{code}
  :- table path/2.
\end{code}
Also, to return
all facts of \co{path} with query-driven inference, additional rules like
the following can be used, employing \co{fail} to continue the search for
each next \co{path} fact until no more can be found and then the last
\co{printPath} succeeds.
\begin{code}
  printPath :- path(X,Y), write(path(X,Y)), fail.
  printPath.
\end{code}

\item {\bf ASP system clingo.} With ground-and-solve, by default all facts
  will be returned. To get back only facts of \co{path}, the following
  additional directive can be used:
\begin{code}
  #show path/2.
\end{code}

\item {\bf Datalog system \souf.} To compile Datalog rules to standalone
  code in C++, which is then compiled and run, additional type
  and\,input-output declarations are\,needed:\!\!
\begin{code}
  .decl edge(x:number, y:number)
  .input edge
  .decl path(x:number, y:number)
  .output path
\end{code}

\end{itemize}
Additionally, Prolog systems like XSB are highly powerful and expressive,
and can be used for direct scripting and timing for performance analysis.
In contrast, ASP and Datalog languages like clingo and \souf have more
limited features and rely on a host language---Python for clingo and C++
for \souf---for application development.

\newcommand{\graphs}{
\begin{table*}[tbp]
  \tlp{\footnotesize}\nottlp{}
  \centering
\tlp{\renewcommand{\arraystretch}{0}}
\begin{tabular}{@{\nottlp{}}l@{~}||@{\nottlp{}}p{\tlp{4}\aaai{}\arxiv{}in}@{~}||@{\nottlp{}\tlp{\!\!}}l@{}}
  Name  & Definition, fully and precisely captured by the set of edges
        & \tlp{BW 2019}\aaai{}\arxiv{}\\\hline\hline
\gname{Cmpl}{n}      & complete graph with vertices \co{1..n},\newline
                       with an edge from every vertex to every vertex\newline
                       \co{\{(i,j):~i in 1..n, j in 1..n\}}
  & \m{K_n}\\\hline
\gname{MaxAcyc}{n}   & maximum acyclic graph with vertices \co{1..n},\newline
                       with an edge from every vertex to every 
                       higher-indexed vertex\newline
                       \co{\{(i,j):~i in 1..n-1, j in i+1..n\}} %
  & \m{T_n}\\\hline
\gname{Cyc}{n}       & cycle going around vertices \co{1..n} in order\newline
                       \co{\{(i,i+1):~i in 1..n-1\} + \{(n,1)\}}
  & \m{C_n}\\\hline
\gname{CycExtra}{n,k}& \gname{Cyc}{n} with \co{k} extra edges from each
                       vertex \co{i} to \co{k} vertices so that these \co{k+1}
                       vertices are evenly spaced on the cycle\newline
                       \co{\{(i,(i-1 + t*n/(k+1)) mod n + 1):~i in 1..n, t in
                       1..k\} + \gname{Cyc}{n}}
  & \m{S_{n,k}}\\\hline
\gname{Path}{n}      & path going through vertices \co{1..n} in order\newline
                       \co{\{(i,i+1):~i in 1..n-1\}}\hfill %
  & \m{P_n}\\\hline
\gname{PathDisj}{n,k}& \co{k} disjoint paths, each going through \co{n}
                       vertices in order\newline
                       \co{\{(i,i+k):~i in 1..(n-1)*k\}}
  & \m{M_{n,k}}\\\hline
\gname{Grid}{n} & Grid of \co{n}-by-\co{n} vertices in order,
                  with edges toward higher-indexed vertices\newline
                  \co{\{(j, j+1):~i in 1..n, j in (i-1)*n+1..i*n-1\} +}\newline
                  \co{\{(j, j+n):~i in 1..n-1, j in (i-1)*n+1..i*n\}}
  & --\\\hline
\gname{BinTree}{h}   & complete binary tree of \co{h} levels of vertices
                       with edges toward the leaves\newline
                       \co{\{(i,2*i):~i in 1..2\m{^{\tt{h-1}}}\} +
                       \{(i,2*i+1):~i in 1..2\m{^{\tt{h-1}}}\}}
  & \m{B_h}\\\hline
\gname{BinTreeRev}{h}& complete binary tree of \co{h} levels of vertices
                       with edges toward the root\newline
                       \co{\{(2*i,i):~i in 1..2\m{^{\tt{h-1}}}\} +
                       \{(2*i+1,i):~i in 1..2\m{^{\tt{h-1}}}\}}
  & \m{V_h}\\\hline
\gname{X}{n,k} & X shaped with an edge from each of \co{n} vertices 
                 to a central vertex \co{n+1}, and an edge from the central
                 vertex to each of \co{k} vertices\newline
                 \co{\{(i,n+1):~i in 1..n\} + \{(n+1,n+1+j):~j in 1..k\}}
  & \m{X_{n,k}}\\\hline
\gname{Y}{n,k} & Y shaped with an edge from each of \co{n} vertices
                 to a central vertex \co{n+1}, and a path going through 
                 \co{k} vertices starting from the central vertex\newline
                 \co{\{(i,n+1):~i in 1..n\} + \{(i,i+1):~i in n+1..n+k-1\}}
  & \m{Y_{n,k}}\\\hline
\gname{W}{n,k} & W shaped subgraphs, formed with two sets of \co{n} vertices
                 each, with an edge from each vertex in the first set to \co{k}
                 vertices in the second set\newline
                 \co{\{(i,n+1 + (i+j-1) mod n):~i in 1..n, j in 1..k\}}
  & \m{W_{n,k}}\\\hline
\end{tabular}
\caption{Kinds of input graphs and their definitions, with their names if
  any in~\citet{brass2019performance}.\Vex{-0}}
\label{tab-graphs}
\end{table*}
} %
\graphs

\mysec{Rule variants and input graph types}
\label{sec-rules-graphs}

We consider subtly different rule variants and a wide variety of graph
types that may affect efficiency of inference methods, drastically.

\mypar{Variants of rules}
It is long known that different variants of rules for solving the same
problem can have drastically different performance, but this has not been
studied precisely in general.  We consider all three well-known variants of
rules for transitive closure:
\begin{enumerate}
\item Left recursion: the recursive rule can be written differently, by
  switching the two predicates, so that the recursive occurrence of
  \co{path} is on the left:
\begin{code}
  path(X,Y) :- path(X,Z), edge(Z,Y).
\end{code}
\item Right recursion: the recursive rule as written in Section
  \ref{sec-sys} has recursive occurrence of \co{path} on the right.
\item Double recursion: the recursive rule can also be written by replacing
  \co{edge} with \co{path}, yielding two recursive occurrences of
  \co{path}:
\begin{code}
  path(X,Y) :- path(X,Z), path(Z,Y).
\end{code}
\end{enumerate}
\myendneg

\mypar{Kinds of input graphs}
We consider and precisely define 12 different kinds of graphs, shown in
Table~\ref{tab-graphs}.  It includes all 11 kinds in
rbench~\cite{brass2019performance}, created to address the issues with
input in OpenRuleBench~\cite{Lia+09open}. We created the last one, grid,
with more rigid and intertwined edges.
Together, they aim to capture graphs of different shapes that may affect
the efficiency of computing the transitive closure very differently.

Note that \co{n} is just one of the parameters, and is not always the
number of vertices in the graph, although it is for 
the first 5 graph types in Table~\ref{tab-calc}.  Table~\ref{tab-calc}
columns 2 and 3 show the precise numbers of vertices and edges.

\mysec{Analysis of optimal time complexities} %
\label{sec-analyze}

To compute 
all 
facts that can be inferred, any inference using bottom-up, group-and-solve,
or top-down methods must consider all combinations of given and inferred
facts that can make all hypotheses of a rule become true.  Each such
combination leads to a rule firing.

Optimality in considering such combinations can be achieved by using the
method of incrementalization with minimum
increment~\cite{\full{}LiuSto09Rules-TOPLAS,Liu13book}.
The method
allows each combination to be considered at most once and in worst-case
\m{O(1)} time---by considering only one
fact at a time and using indexing to find each new combination with the
fact in 
\m{O(1)}
time.

\newcommand{\analysis}{
\newcommand{\asleft}{same as left}
\begin{table*}[t]
\tlp{\footnotesize}\nottlp{}
  \centering
\tlp{\renewcommand{\arraystretch}{0}}
\begin{tabular}{@{\nottlp{}}l@{~}||@{\nottlp{}}l@{~}|@{\nottlp{}}l@{~}|@{\nottlp{}}p{\tlp{0.75}\aaai{}\arxiv{}in}@{~}||@{\nottlp{}}p{\tlp{1.}\aaai{}\arxiv{}in}@{~}|@{\nottlp{}}p{\tlp{1}\aaai{}\arxiv{}in}@{~}}
Graphs                & \#vertex  & \#edge      & \#path
  & Left recursion and\newline right recursion
  & Double recursion
\\\hline\hline
\gname{Cmpl}{n}       & \m{n}     & \m{n^2}     & \m{n^2} 
  & \m{n^3}
  & \asleft \\\hline
\gname{MaxAcyc}{n}    & \m{n}     & \m{\frac{1}{2}n(n-1)}& \m{\frac{1}{2}n(n-1)}
  & \m{\frac{1}{6}n(n-1)(n-2)}
  & \asleft \\\hline
\gname{Cyc}{n}        & \m{n}     & \m{n}       & \m{n^2} 
  & \m{n^2}
  & \m{n^3} \\\hline
\gname{CycExtra}{n,k} & \m{n}     & \m{n+nk}    & \m{n^2} 
  & \m{n^2+n^2k}
  & \m{n^3} \\\hline
\gname{Path}{n}       & \m{n}     & \m{n-1}     & \m{\frac{1}{2}n(n-1)}
  & \m{\frac{1}{2}(n-1)(n-2)}
  & \m{\frac{1}{6}n(n-1)(n-2)}\\\hline
\gname{PathDisj}{n,k} & \m{n k}   & \m{(n-1)k}  & \m{\frac{1}{2}n(n-1)k}
  & \m{\frac{1}{2}(n-1)(n-2)k}
  & \m{\frac{1}{6}n(n-1)(n-2)k}\\\hline
\gname{Grid}{n}       & \m{n^2}   & \m{2n(n-1)} & \m{\frac{1}{4}n^2(n+3)(n-1)}
  & \m{\frac{1}{2}n(n^3-5n+4)}
  & \m{\frac{1}{36}(n^3 + 7 n^2 + 2 n - 22) n^2 (n-1)} \\\hline
\gname{BinTree}{h}    & \m{2^h-1} & \m{2^h-2}   & \m{2^h (h-2) + 2}
  & \m{2^h(h-3) + 4}
  & \m{2^{h - 1} (h^2 - 5 h + 8) - 4}\\\hline
\gname{BinTreeRev}{n} & \m{2^h-1} & \m{2^h-2}   & \m{2^h (h-2) + 2}
  & \m{2^h(h-3) + 4}
  & \m{2^{h - 1} (h^2 - 5 h + 8) - 4}\\\hline
\gname{X}{n,k}        & \m{n+k+1} & \m{n+k}     & \m{n+k+nk}
  & \m{nk}
  & \asleft \\\hline
\gname{Y}{n,k}        & \m{n+k}   & \m{n+k-1}   & \m{\frac{1}{2} (2n+k-1) k}
  & \m{\frac{1}{2} (2n+k-2)(k-1)}
  & \m{\frac{1}{6} (3n + k - 2) k (k - 1)} \\\hline
\gname{W}{n,k}        & \m{2n}    & \m{n k}    & \m{n k}
  & \m{0}
  & \asleft \\\hline
\end{tabular}
\caption{Optimal number of combinations for all 3 recursion variants and all 12 graph types.\Vex{-0}}
\label{tab-calc}
\end{table*}
} %
\analysis
\mysubsec{Optimal number of combinations}
\label{sec-optimal}
For computing all \co{path} facts, for all three rule variants, the first
rule is fired once for each \co{edge} fact, totaling to be \co{\#edge}, the
number of edges.
For the second, recursive rule,
we calculate the exact number of combinations for each graph type and each
rule variant.
The calculation, for each graph type and each rule variant, first forms a
precise formula for the exact counting and summing, and then simplifies the
formula to a closed form using Mathematica.
Details of the calculation are in \tlp{an appendix of~\citet{Liu+25RuleInfEfficiency-arxiv}}\arxiv{}.

Table~\ref{tab-calc} shows the resulting precise complexity formulas in
closed forms for the recursive rule.
The particular closed forms in Table~\ref{tab-calc} were selected and
rearranged manually to have a consistent factor representation across all
rule variants and graph types.

Despite all the differences among the kinds of graphs listed, there are a
number of interesting equalities to observe.  For example, for \co{Cmpl}
graphs, the number of combinations is the same for all different
recursions, and this is also the case for \co{MaxAcyc}, \co{X}, and \co{W} graphs.

It is particularly interesting to see that for all graph types,
left and right recursions lead to the same number of combinations.
This is easy to see for graphs with a kind of symmetry between paths going
into an edge (left recursion) and paths coming out (right recursion),
which include the first 7 graph types.
It is also easy to see for \gname{X}{} and \gname{W}{} graphs, which are
the simplest cases with only paths of lengths 2 and 1, respectively.

However, \gname{BinTree}{h} and \gname{BinTreeRev}{h} do not have such
symmetry, similar to \gname{Y}{n,k} not having such symmetry.  It still
happens to be that even for left or right recursion alone,
\gname{BinTree}{h} and \gname{BinTreeRev}{h} have the same number of
combinations despite they are from very different summation formulas.
Consider the right recursion form \co{edge(X,Y), path(Y,Z)}.
\begin{itemize}
\item

For \gname{BinTree}{h}, consider \co{i-1} levels of edges from root to
leaves, with \co{i} from \co{1} to \co{h-1}.
At the \co{i}th level of edges, there are \co{2\m{^{\tt i}}} edges.
For each such edge \co{(u,v)}, there is a path from \co{v} to each
vertex in the subtree rooted at \co{v} except for \co{v}.  The subtree has
\co{2\m{^{\tt h-i}}-1} vertices, and the connecting vertex itself should 
also be subtracted.  Thus, the number of combinations for
\co{edge(X,Y), path(Y,Z)} is
\(
\sum_{i = 1..h-1} 2^i * (2^{h-i}-2)
\).

\item 

For \gname{BinTreeRev}{h}, consider \co{i-1} levels of edges from leaves to
root, with \co{i} from \co{h} to \co{2}. %
At the \co{i}th level of edges, there are \co{2\m{^{\tt i-1}}} edges.
For each such edge \co{(u,v)}, there is a path from \co{v} to each vertex
above it in the path to root.  There are \co{i-2} such vertices.
Thus, the number of combinations for \co{edge(X,Y), path(Y,Z)} is
\(
\sum_{i = h..2} 2^{i-1} * (i-2)
\).
\end{itemize}
Both actually simplify to the same closed form as in Table~\ref{tab-calc}.
Because of this same closed form, left and right recursions for both kinds
of trees have the same closed form too.
When we found that left and right recursions for \co{Y} graphs
also have the same closed form, it became curious, and we
prove the following general result for directed trees;
the result actually applies to \gname{Path}{n}, \gname{PathDisj}{n,k},
and the last 5 graph types.

We prove that for any directed tree, left and right recursions have the
same number of combinations.  Precisely, let \co{vert} be the set of
vertices in the given \co{edge} relation, then
\[
\sum_{\co{v} \in \co{vert}} \#\{\co{u:~path(u,v)}\} \times \#\{\co{w:~edge(v,w)}\}
=
\sum_{\co{v} \in \co{vert}} \#\{\co{u:~edge(u,v)}\} \times \#\{\co{w:~path(v,w)}\}
\]

For any directed tree, there is at most one path between any two vertices.
Consider any matching \co{path(u,v)} and \co{edge(v,w)} from left
recursion. Let \co{edge(u,v1)} be the unique edge on the path from \co{u}
to \co{v}.  Then there is the unique \co{path(v1,v)} joining with
\co{edge(v,w)} giving the unique \co{path(v1,w)}.  Now the matching
\co{edge(u,v1)} and \co{path(v1,w)} must be counted in right recursion.
Thus each combination \co{path(u,v)} and \co{edge(v,w)} from left recursion
corresponds to a unique combination \co{edge(u,v1)} and \co{path(v,w)} from
right recursion.
Symmetrically, each combination from right recursion corresponds to a
unique combination from left recursion.  Thus, the equation holds.

Note that the equation does not hold for all graphs.  An example is a graph
with edges \co{\{(1,2),\,(2,1),\,(1,3),\,(2,3)\}}. The combinations
\co{path(1,1),\,edge(1,2)} and \co{path(1,1),\,edge(1,3)} are not captured
in right recursion, because there is no edge \co{(1,1)} or \co{(2,2)}.

\mysubsec{Time complexities of different inference methods}
\label{sec-difftime}

Despite that optimal complexities are achieved with
incrementalization-based
method~\cite{LiuSto03Rules-PPDP,LiuSto09Rules-TOPLAS} and confirmed
experimentally~\cite{LiuSto09Rules-TOPLAS}, implementations using other
inference methods have different other cost issues.
\begin{description}

\item {\bf Query-driven inference with tabling} has separate overhead from
  general mechanisms for table maintenance and lookup. However, this is
  minimal, but the following is a significantly larger cost issue.

  The inference considers hypotheses of a rule from left to right, so the
  order of the two predicates matters. 
  Contrary to common belief, left recursion is the most efficient for table
  lookup,
  as analyzed precisely in~\citet{TekLiu10RuleQuery-PPDP};
  the key idea is that, with left recursion, the search for \co{path(X,Y)}
  looks up \co{path(X,Z)} in the same table due to the same value of
  \co{X}.
  With right recursion, it needs to look up \co{path(Z,Y)} in a different
  table for each different value of \co{Z} from \co{edge(X,Z)} on the left;
  this is also the case with double recursion except that the \co{Z} is
  from \co{path(X,Z)}, which could be asymptotically larger than from
  \co{edge(X,Z)}
  depending on the input graph. %

\item {\bf Ground-and-solve with optimized grounding and solving} generally
  has a separate cost tracking ground rules rather than directly returning
  the derived facts.
  This is not minimal but %
  relatively small.  For recursive rules with
  no negation, grounding does iterative evaluation on ground rules and
  computes all resulting facts.

  A larger cost issue is similar to the issue with recursion variants for
  query-driven inference above, because grounding also considers hypotheses
  in a rule from left to right~\cite{kaufmann2016grounding}.  Additionally,
  despite extensive optimizations,
  iterative evaluation on ground rules uses semi-naive 
  evaluation~\cite{kaufmann2016grounding}, another
  larger cost issue as described for fact-driven inference below.

\item {\bf Fact-driven inference with semi-naive evaluation} has redundant
  computations and overhead from considering all new facts
  in each iteration.
  This is because it goes through an extra set, and there could be repeated
  computations within the processing of a set instead of one
  element~\cite{CaiPai88}, unlike incrementalization with minimum
  increment~\cite{\full{}LiuSto09Rules-TOPLAS,Liu13book}.
  It could be asymptotically worse than optimal without prudent incremental
  processing of the extra set.

  However, in contrast to the two methods above, inference driven by newly
  inferred facts of a predicate, e.g., \co{path}, can have the same
  efficiency regardless of whether the predicate is in a left or right
  recursion to join with another predicate.

  Finally, fact-driven inference often generates standalone code from rules,
  which has a separate cost of compilation, but this is needed only once
  for each program.
\end{description}

Note that the precise formulas in Table~\ref{tab-calc} support comparisons
of not only asymptotics but also constants.  For example, for
\co{Cmpl} graphs, all three recursion variants have exactly the same
number; in comparison, for \co{MaxAcyc} with about 1/2 the size of input
(\co{edge}) and output (\co{path}), left and right recursions have about
1/6 of the number of combinations, whereas double recursion has the same
number.
Section~\ref{sec-expe} shows through experiments how difference inference
methods and systems match the analytical results in this section.

\tlp{
\def\plotdir{plots}
\def\ploteqdir{plots/k=n}
\def\plotBAdir{plots/realworld/barabasi_albert}
}
\aaai{}
\arxiv{}
\newcommand{\plotfigdir}[4]{
\begin{figure*}[t]
  \centering
  \begin{subfigure}[b]{\tlp{0.325}\aaai{}\arxiv{}\textwidth}
    \includegraphics[width=\linewidth,height=.73\linewidth]{#4/lr/#1.pdf}\Vex{-1}
    \caption{Left recursion}
  \end{subfigure}
  \hfill
  \begin{subfigure}[b]{\tlp{0.325}\aaai{}\arxiv{}\textwidth}
    \includegraphics[width=\linewidth,height=.73\linewidth]{#4/rr/#1.pdf}\Vex{-1}
    \caption{Right recursion}
  \end{subfigure}
  \hfill
  \begin{subfigure}[b]{\tlp{0.325}\aaai{}\arxiv{}\textwidth}
    \includegraphics[width=\linewidth,height=.73\linewidth]{#4/dr/#1.pdf}\Vex{-1}
    \caption{Double recursion}
  \end{subfigure}
  \Vex{-1.75} 

  \caption{Running times on \co{#1} graphs, up to #3 vertices, #2 edges.\Vex{-0}}
  \label{fig-#1}
\end{figure*}
}
\newcommand{\plotfig}[3]{\plotfigdir{#1}{#2}{#3}{\plotdir}}
\newcommand{\ploteqfig}[3]{\plotfigdir{#1}{#2}{#3}{\ploteqdir}}
\newcommand{\plotBAfig}[3]{\plotfigdir{#1}{#2}{#3}{\plotBAdir}}

\notsep{}

\sepexpe{
\mysec{Key findings from experimental results and analysis}
\label{sec-expe}

We designed and performed detailed experiments with all three rule systems
XSB, clingo, and \souf, all three rule variants, and all 12 input graph
types.  The programs, variants of each, and graphs are exactly as described in
Sections 2.2 and 3.
For example, Figs.~\notsep{}\sepexpe{1} 
and~\notsep{}\sepexpe{2} %
show running times on \co{Cmpl} and \co{Path} %
graphs for all recursion types in all three rule systems.
Details of the experiments and results are in \aaai{}\tlp{an appendix in~\citet{Liu+25RuleInfEfficiency-arxiv}}\arxiv{}.%
\plotfig{Cmpl}{1000000}{1000}%
\plotfig{Path}{999}{1000}%
\ploteqfig{W}{1000000 (\m{k=n=1000})}{2000}%
}%
Key findings from comparing and confirming the\notsep{}\sepexpe{ analyzed results} with the
experimental results\notsep{} are as follows:

\begin{enumerate}
\item Overall, precise calculation of optimal complexity and dedicated
  analysis of different inference and optimization methods in
  Section~\ref{sec-analyze} help tremendously in predicting their
  performances and understanding the differences, for both asymptotic
  trends and constant factors, across not only the inference and
  optimization methods but also rule variants and graph types.

\item Which inference method and system to use depends on the unique
  features and powers the method and system provide, including the
  advantages summarized in Table~\ref{tab-sys}. Whichever method and system
  are used, the rule variants and input graph types can affect the
  performance much more, as described below.

\item Efficient recursion types depend on the inference method and system.
  Except for an XSB performance bug discovered for the simplest case (left
  recursion on \gname{W}{n,k}), described below, and relatively small XSB
  and \souf variations on the next simplest case
  (left recursion on \gname{X}{n,k}), (1) for XSB and clingo, left
  recursion is fastest;
  (2) for \souf, left and right recursions are about the same, faster than
  double recursion;
  and (3) overall, left recursion is fastest, double recursion can be
  asymptotically slower, all as analyzed in Section~\ref{sec-analyze}.

\item On all graphs except for left recursion on \gname{W}{n,k} due to the
  performance bug, XSB, with query-driven inference with tabling, is the
  fastest, by several times or even an order of magnitude.
  clingo is the next fastest except for right recursion for
  \co{Cmpl} (e.g., Fig.~\notsep{}\sepexpe{1}) and
  \co{MaxAcyc} where \souf is faster.
  \souf is fastest in reading data and writing results (e.g.,
  Fig.~\notsep{}\sepexpe{3}), even if it has the extra
  compilation time.

  We discovered the XSB performance bug when running on \gname{W}{n,k} with
  increasing \co{k} = \co{n} (Fig.~\notsep{}\sepexpe{3}). This
  is the simplest case of all graph types, where \co{path} is exactly
  \co{edge} with no extra paths at all, but XSB on left recursion is
  drastically slower than analyzed.
  XSB team confirmed this performance bug and analyzed that it is due to
  hash collisions in this case.

\item Besides fixing the XSB bug, XSB and Clingo could be improved by
  automatically converting right recursion to left
  recursion~\cite{Tek+08RulePE-AMAST}.
  Clingo's right and double recursions could perhaps be improved, possibly
  with better incremental processing
  for later hypotheses in a rule.
  \souf has opportunities to become faster in general, for all recursion
  types\full{}, because generated C++ code can be
  specialized to be much faster than general evaluator
  code~\cite{RotLiu07Retrieval-PEPM}.
\end{enumerate}
\plotBAfig{BA}{200000}{100000}

We also used well-known graph generators designed to simulate real-world
networks and experimented with all three recursion types using all three
rule systems.  Fig.~\ref{fig-BA} shows the running times on directed
Barab\'asi-Albert (BA)
graphs~\cite{%
albert2002statistical}, the most famous
model for explaining the scale-free nature of the Internet, World Wide Web,
social media, and citation networks. 
The graphs are generated using
the widely-used library NetworkX with preferential attachment parameter value 2.

The results in Fig.~\ref{fig-BA}
are consistent with
everything we have analyzed, in terms of relative performances of all %
recursion types and all %
systems, including a similar performance
anomaly in left recursion in XSB as \co{W} graphs in Fig.~3.  We think the
anomaly is due to the same reason, because BA graphs are known to have %
ultra-small path lengths, with average length growing logarithmically
with the number of vertices, whereas \co{W} graphs have only paths of length
1, so the anomaly for BA graphs is not as serious as for \co{W} graphs.

\mysec{Related work and conclusion}
\label{sec-related}

There has been extensive study of efficiency of inference methods and
systems, from performance benchmarking to complexity analysis.
We discuss closely related works.

Prolog has had benchmarks for comparing performance of different
implementations, e.g.,~\cite{cmu85prologbench,unipr01clpbench}.
Some are focused on a special class of problems, e.g., interpreters written
in Prolog~\cite{korner2020performance}.  Some are for evaluating the
performance of a particular implementation, e.g., SWI
Prolog~\cite{swi26bench}.

There are also works that evaluate queries in more general rule engines and
database systems, e.g., the LUBM benchmark~\cite{guo2005lubm} and its
extensions~\cite{ma2006towards,singh2020owl2bench} for OWL
on a range of systems,
and evaluations including SQL with
rules~\cite{brass2019new,brass2019performance}.
In particular, %
rbench~\cite{brass2019performance} considers 11 types of
graphs, and we use all of them.

OpenRuleBench~\cite{Lia+09open} was extensively constructed for
comprehensively evaluating a wide range of systems on a diverse set of
problems.
Benchmarking~\cite{Liu+23RuleLangBench-ICLP} for
Alda~\cite{Liu+23RuleLangInteg-TPLP}
used all OpenRuleBench benchmarks and more to evaluate an
extension of DistAlgo~\cite{Liu+17DistPL-TOPLAS}/Python with Datalog rules
that implements queries using XSB,
and to compare with programs written purely in Python,
DistAlgo, and XSB.

More studies of methods and systems compare on some chosen workloads\full{}, e.g.,
\full{}Flan~\cite{abeysinghe2024flan} on adding and compiling Datalog to Scala
compares with \souf and two Datalog systems embedded in Rust, Crepe and
Ascent,
on program analysis;
BMLP~\cite{ai2024boolean} on using Boolean matrix operations for rules
compares with \souf, clingo, B-Prolog, and SWI-Prolog on rules for graphs.

Datalog systems are increasingly
studied~\cite{maier18hist-wbook,vianu2021datalog,ketsman2022modern},
e.g.,~\citet{ketsman2022modern} focuses on 9 Datalog systems 
including \souf, among 17 Datalog engines and more mentioned,
and discusses the extensively-used semi-naive evaluation %
and more.
There are also dedicated works on evaluating their performances,
e.g.,~\cite{fan2022towards,qureshi2024evaluating}.  In
particular,~\citet{fan2022towards} studies semi-naive execution profile,
running time, CPU utilization, and memory consumption on
transitive closure and four other closely related problems on binary
relations.

None of the existing works analyzes and compares together all three widely
different inference methods and their optimizations that make large
performance impacts.
Also, none of them analyzes different rule variants for the same problem
and their significant performance differences.
Furthermore, none of them studies precise optimal complexities, or performs
detailed measurements from compiling and loading
programs to writing query results, let alone confirming measurements
against the complexities across inference methods, rule
variants, and input graph types.

McAllester and Ganzinger~\cite{mcallester99,%
  gan02logical} analyze complexities of bottom-up evaluation using
prefix firing counting.
Liu and Stoller~\cite{LiuSto03Rules-PPDP,LiuSto09Rules-TOPLAS}
develop systematic incrementalization to automatically transform Datalog
rules into efficient implementations with precise time and space
guarantees for optimality and trade-offs.
Tekle and Liu~\cite{TekLiu10RuleQuery-PPDP,TekLiu11RuleQueryBeat-SIGMOD}
precisely analyze complexities of top-down evaluation for demand-driven
queries and develop demand transformations for bottom-up evaluation,
improving over magic-set transformations exponentially in program size.

These studies of complexity analysis have been essential for the optimal
complexity calculation in this paper, and for precisely understanding the
efficiency of top-down query-driven evaluation with tabling vs.\ bottom-up
fact-driven evaluation.  Their extensions to handle unrestricted negation,
quantification, and aggregation, following a powerful unified semantics for
them~\cite{LiuSto20Founded-JLC,LiuSto22RuleAgg-JLC}, will be essential for
critical applications that need those features.

\mypar{Conclusion, limitations, and future works}
We have presented a systematic study of efficiency of analysis of
transitive relations using different inference methods for different rule
variants and input relationship graphs.  Despite the compelling results for
better understanding and utilizing different rule systems, this work has
several limitations, leaving many avenues for future work.

First, in the experiments, the degree of superlinearity and constant costs
are not analyzed systematically; analyzing them systematically can help
better understand and optimize rule systems.

Second, the %
study is only about time efficiency; %
space~\cite{LiuSto09Rules-TOPLAS,TekLiu10RuleQuery-PPDP,TekLiu11RuleQueryBeat-SIGMOD},
including cache effect on large data,
and other measures~\cite{fan2022towards} are important too.
Also important are additional system-specific aspects that affect
performance, e.g., XSB has subsumptive tabling (which is faster in some
cases than default tabling)~\cite{swift2012xsb},
indexing directives, many input/output functions (for efficiency in
different cases), and more~\cite{xsb22}.

Third, comparative analysis of different methods for demand-driven queries,
not inferring all facts, needs to be studied.
Demand
transformations~\cite{TekLiu10RuleQuery-PPDP,TekLiu11RuleQueryBeat-SIGMOD}
would help clingo and \souf; otherwise, goal-directed queries in XSB would
perform drastically better.  Magic sets in the intelligent grounder of
DLV~\cite{calimeri2017dlv} and in \souf~\cite{souffle26github} are
important for practical efficiency.  Transforming into best recursion
variants~\cite{Tek+08RulePE-AMAST} will also help significantly, as well as
employing modular approaches, e.g,~\cite{hu2022modular}.

Finally, our comparative analysis is only for analyzing transitive
relations;
many more analysis problems can be added,
especially those with negation and aggregation in existing rule
systems~\cite{SagSW94xsb,jordan2016souffle,gebser2019multi} and even
quantification, all
unrestricted\full{}.

\Vex{1.5}
\mypar{Acknowledgment}

We thank Prof.\ David Warren for
discussions about XSB performance and for confirming and analyzing the
XSB performance bug we discovered.  We also thank anonymous reviewers
for their very helpful comments, especially for giving the small example 
at the end of Section~4.1.
This work was supported in part by NSF under grant CCF-1954837. %
\myendneg

{
\arxiv{} 
\def\usebib{
\def\bibdir{../../../bib}   %
{
\bibliography{\bibdir/strings,\bibdir/liu,\bibdir/IC,\bibdir/PT,\bibdir/PA,\bibdir/Lang,\bibdir/Algo,\bibdir/DB,\bibdir/AI,\bibdir/Sec,\bibdir/Sys,\bibdir/Perform,\bibdir/SE,\bibdir/Vis,\bibdir/misc,\bibdir/crossref} %

\begin{thebibliography}{}

\bibitem[Abeysinghe et~al., 2024]{abeysinghe2024flan}
{\sc Abeysinghe, S.}, {\sc Xhebraj, A.}, {\sc and} {\sc Rompf, T.} 2024.
\newblock Flan: An expressive and efficient {Datalog} compiler for program
  analysis.
\newblock {\em Proceedings of the ACM on Programming Languages}, {\it 8}, POPL,
  2577--2609.

\bibitem[Ai and Muggleton, 2024]{ai2024boolean}
{\sc Ai, L.} {\sc and} {\sc Muggleton, S.~H.} 2024.
\newblock Boolean matrix logic programming.
\newblock {\em Computing Research Repository}, {\it arXiv:2408.10369}.

\bibitem[Albert and Barab{\'a}si, 2002]{albert2002statistical}
{\sc Albert, R.} {\sc and} {\sc Barab{\'a}si, A.-L.} 2002.
\newblock Statistical mechanics of complex networks.
\newblock {\em Reviews of modern physics}, {\it 74}, 1, 47.

\bibitem[Bancilhon, 1986]{bancilhon1986naive}
{\sc Bancilhon, F.}
\newblock Naive evaluation of recursively defined relations.
\newblock In {\em On Knowledge Base Management Systems: Integrating Artificial
  Intelligence and Database Technologies} 1986, pp. 165--178. Springer.

\bibitem[Bellomarini et~al., 2018]{bellomarini2018vadalog}
{\sc Bellomarini, L.}, {\sc Sallinger, E.}, {\sc and} {\sc Gottlob, G.} 2018.
\newblock The {Vadalog} system: {Datalog}-based reasoning for knowledge graphs.
\newblock {\em Proceedings of the VLDB Endowment}, {\it 11}, 9.

\bibitem[Brass and Wenzel, 2019a]{brass2019new}
{\sc Brass, S.} {\sc and} {\sc Wenzel, M.}
\newblock A new benchmark database and an analysis of transitive closure
  runtimes.
\newblock In {\em Post-Proceedings of the 32nd Workshop on (Constraint) Logic
  Programming} 2019a, pp. 3--18.
\newblock \url{https://wi.hwtk.de/WLP2018/Papers/WLP_2018_paper_1.pdf}.

\bibitem[Brass and Wenzel, 2019b]{brass2019performance}
{\sc Brass, S.} {\sc and} {\sc Wenzel, M.}
\newblock Performance analysis and comparison of deductive systems and {SQL}
  databases.
\newblock In {\em Proceedings of the 3rd International Workshop on the
  Resurgence of Datalog in Academia and Industry} 2019b, pp. 27--38.
  CEUR-WS.org.

\bibitem[Cai and Paige, 1988]{CaiPai88}
{\sc Cai, J.} {\sc and} {\sc Paige, R.} 1988.
\newblock Program derivation by fixed point computation.
\newblock {\em Science of Computer Programming}, {\it 11}, 197--261.

\bibitem[Calimeri et~al., 2017]{calimeri2017dlv}
{\sc Calimeri, F.}, {\sc Fusc{\`a}, D.}, {\sc Perri, S.}, {\sc and} {\sc
  Zangari, J.} 2017.
\newblock {I-DLV}: {The} new intelligent grounder of {DLV}.
\newblock {\em Intelligenza Artificiale}, {\it 11}, 1, 5--20.

\bibitem[Chen and Warren, 1996]{CheWar96}
{\sc Chen, W.} {\sc and} {\sc Warren, D.~S.} 1996.
\newblock Tabled evaluation with delaying for general logic programs.
\newblock {\em Journal of the ACM}, {\it 43}, 1, 20--74.

\bibitem[CHINA, 2001]{unipr01clpbench}
CHINA 2001.
\newblock {The CHINA Benchmark Suite}.
\newblock \url{http://www.cs.unipr.it/China/Benchmarks}.
\newblock Accessed Apr.\ 22, 2025.

\bibitem[clingo, 2026]{clingo26}
clingo 2026.
\newblock {clingo and gringo}.
\newblock \url{https://potassco.org/clingo}.
\newblock Accessed Feb.\ 27, 2026.

\bibitem[{CMU AI Repository}, 1985]{cmu85prologbench}
{CMU AI Repository} 1985.
\newblock {Prolog Benchmarking Suites}.
\newblock
  \url{http://www.cs.cmu.edu/afs/cs/project/ai-repository/ai/lang/prolog/code/bench/0.html}.
\newblock Accessed Feb.\ 27, 2026.

\bibitem[Fan et~al., 2022]{fan2022towards}
{\sc Fan, Z.}, {\sc Mallireddy, S.}, {\sc and} {\sc Koutris, P.}
\newblock Towards better understanding of the performance and design of
  {Datalog} systems.
\newblock In {\em Datalog 2.0: 4th International Workshop on the Resurgence of
  Datalog in Academia and Industry} 2022, pp. 166--180.

\bibitem[Flores-Montoya and Schulte, 2020]{flores2020datalog}
{\sc Flores-Montoya, A.} {\sc and} {\sc Schulte, E.}
\newblock Datalog disassembly.
\newblock In {\em 29th USENIX Security Symposium (USENIX Security 20)} 2020,
  pp. 1075--1092.

\bibitem[Ganzinger and McAllester, 2002]{gan02logical}
{\sc Ganzinger, H.} {\sc and} {\sc McAllester, D.~A.}
\newblock Logical algorithms.
\newblock In {\em Proceedings of the 18th International Conference on Logic
  Programming} 2002, pp. 209--223. Springer.

\bibitem[Gebser et~al., 2012]{GebKKS12book}
{\sc Gebser, M.}, {\sc Kaminski, R.}, {\sc Kaufmann, B.}, {\sc and} {\sc
  Schaub, T.} 2012.
\newblock {\em Answer Set Solving in Practice}.
\newblock Synthesis Lectures on Artificial Intelligence and Machine Learning.
  Morgan \& Claypool.

\bibitem[Gebser et~al., 2019]{gebser2019multi}
{\sc Gebser, M.}, {\sc Kaminski, R.}, {\sc Kaufmann, B.}, {\sc and} {\sc
  Schaub, T.} 2019.
\newblock Multi-shot {ASP} solving with clingo.
\newblock {\em Theory and Practice of Logic Programming}, {\it 19}, 1, 27--82.

\bibitem[Gr{\"a}del, 1991]{gradel1991transitive}
{\sc Gr{\"a}del, E.}
\newblock On transitive closure logic.
\newblock In {\em International Workshop on Computer Science Logic} 1991, pp.
  149--163. Springer.

\bibitem[Guo et~al., 2005]{guo2005lubm}
{\sc Guo, Y.}, {\sc Pan, Z.}, {\sc and} {\sc Heflin, J.} 2005.
\newblock {LUBM}: A benchmark for {OWL} knowledge base systems.
\newblock {\em Journal of Web Semantics}, {\it 3}, 2-3, 158--182.

\bibitem[Hogan et~al., 2021]{hogan2021knowledge}
{\sc Hogan, A.}, {\sc Blomqvist, E.}, {\sc Cochez, M.}, {\sc d’Amato, C.},
  {\sc Melo, G.~D.}, {\sc Gutierrez, C.}, {\sc Kirrane, S.}, {\sc Gayo, J.
  E.~L.}, {\sc Navigli, R.}, {\sc Neumaier, S.}, {\sc and} {\sc others} 2021.
\newblock Knowledge graphs.
\newblock {\em ACM Computing Surveys}, {\it 54}, 4, 1--37.

\bibitem[Hristova et~al., 2007]{Hri+07SPKI-PPDP}
{\sc Hristova, K.}, {\sc Tekle, K.~T.}, {\sc and} {\sc Liu, Y.~A.}
\newblock Efficient trust management policy analysis from rules.
\newblock In {\em Proceedings of the 9th ACM SIGPLAN International Conference
  on Principles and Practice of Declarative Programming} 2007, pp. 211--220.
  ACM Press.

\bibitem[Hu et~al., 2022]{hu2022modular}
{\sc Hu, P.}, {\sc Motik, B.}, {\sc and} {\sc Horrocks, I.} 2022.
\newblock Modular materialisation of datalog programs.
\newblock {\em Artificial Intelligence}, {\it 308}, 103726.

\bibitem[Jordan et~al., 2016]{jordan2016souffle}
{\sc Jordan, H.}, {\sc Scholz, B.}, {\sc and} {\sc Suboti{\'c}, P.}
\newblock Souffl{\'e}: {On} synthesis of program analyzers.
\newblock In {\em Proceedings of the Intl.\ Conference on Computer Aided
  Verification} 2016, pp. 422--430. Springer.

\bibitem[Kaufmann et~al., 2016]{kaufmann2016grounding}
{\sc Kaufmann, B.}, {\sc Leone, N.}, {\sc Perri, S.}, {\sc and} {\sc Schaub,
  T.} 2016.
\newblock Grounding and solving in answer set programming.
\newblock {\em AI Magazine}, {\it 37}, 3, 25--32.

\bibitem[Ketsman et~al., 2022]{ketsman2022modern}
{\sc Ketsman, B.}, {\sc Koutris, P.}, {\sc and} {\sc others} 2022.
\newblock Modern {Datalog} engines.
\newblock {\em Foundations and Trends{\textregistered} in Databases}, {\it 12},
  1, 1--68.

\bibitem[K{\"o}rner et~al., 2020]{korner2020performance}
{\sc K{\"o}rner, P.}, {\sc Schneider, D.}, {\sc and} {\sc Leuschel, M.}
\newblock On the performance of bytecode interpreters in {Prolog}.
\newblock In {\em International Workshop on Functional and Constraint Logic
  Programming} 2020, pp. 41--56. Springer.

\bibitem[Li and Mitchell, 2003]{LiMit03}
{\sc Li, N.} {\sc and} {\sc Mitchell, J.~C.}
\newblock Datalog with constraints: {A} foundation for trust management
  languages.
\newblock In {\em Proceedings of the 5th International Symposium on Practical
  Aspects of Declarative Languages} 2003, pp. 58--73. Springer.

\bibitem[Liang et~al., 2009]{Lia+09open}
{\sc Liang, S.}, {\sc Fodor, P.}, {\sc Wan, H.}, {\sc and} {\sc Kifer, M.}
\newblock {OpenRuleBench}: {A}n analysis of the performance of rule engines.
\newblock In {\em Proceedings of the 18th International Conference on World
  Wide Web} 2009, pp. 601--610. ACM Press.

\bibitem[Liu, 2013]{Liu13book}
{\sc Liu, Y.~A.} 2013.
\newblock {\em Systematic Program Design: {F}rom Clarity to Efficiency}.
\newblock Cambridge University Press.

\bibitem[Liu et~al., 2026]{Liu+25RuleInfEfficiency-arxiv}
{\sc Liu, Y.~A.}, {\sc Idogun, J.}, {\sc Stoller, S.~D.}, {\sc and} {\sc Tong,
  Y.} 2025 (revised May 2026).
\newblock Efficiency of analysis of transitive relations using query-driven,
  ground-and-solve, and fact-driven inference.
\newblock {\em Computing Research Repository}, {\it arXiv:2504.21291 [cs.DB]}.

\bibitem[Liu and Stoller, 2003]{LiuSto03Rules-PPDP}
{\sc Liu, Y.~A.} {\sc and} {\sc Stoller, S.~D.}
\newblock From {Datalog} rules to efficient programs with time and space
  guarantees.
\newblock In {\em Proceedings of the 5th ACM SIGPLAN International Conference
  on Principles and Practice of Declarative Programming} 2003, pp. 172--183.
  ACM Press.

\bibitem[Liu and Stoller, 2009]{LiuSto09Rules-TOPLAS}
{\sc Liu, Y.~A.} {\sc and} {\sc Stoller, S.~D.} 2009.
\newblock From {Datalog} rules to efficient programs with time and space
  guarantees.
\newblock {\em ACM Transactions on Programming Languages and Systems}, {\it
  31}, 6, 1--38.

\bibitem[Liu and Stoller, 2020]{LiuSto20Founded-JLC}
{\sc Liu, Y.~A.} {\sc and} {\sc Stoller, S.~D.} 2020.
\newblock Founded semantics and constraint semantics of logic rules.
\newblock {\em Journal of Logic and Computation}, {\it 30}, 8, 1609--1638.

\bibitem[Liu and Stoller, 2022]{LiuSto22RuleAgg-JLC}
{\sc Liu, Y.~A.} {\sc and} {\sc Stoller, S.~D.} 2022.
\newblock Recursive rules with aggregation: A simple unified semantics.
\newblock {\em Journal of Logic and Computation}, {\it 32}, 8, 1659--1693.

\bibitem[Liu et~al., 2017]{Liu+17DistPL-TOPLAS}
{\sc Liu, Y.~A.}, {\sc Stoller, S.~D.}, {\sc and} {\sc Lin, B.} 2017.
\newblock From clarity to efficiency for distributed algorithms.
\newblock {\em ACM Transactions on Programming Languages and Systems}, {\it
  39}, 3, 12:1--12:41.

\bibitem[Liu et~al., 2023a]{Liu+23RuleLangInteg-TPLP}
{\sc Liu, Y.~A.}, {\sc Stoller, S.~D.}, {\sc Tong, Y.}, {\sc and} {\sc Lin, B.}
  2023a.
\newblock Integrating logic rules with everything else, seamlessly.
\newblock {\em Theory and Practice of Logic Programming}, {\it 23}a, 4,
  678--695.

\bibitem[Liu et~al., 2022]{Liu+22RuleLang-arxiv}
{\sc Liu, Y.~A.}, {\sc Stoller, S.~D.}, {\sc Tong, Y.}, {\sc Lin, B.}, {\sc
  and} {\sc Tekle, K.~T.} 2022.
\newblock Programming with rules and everything else, seamlessly.
\newblock {\em Computing Research Repository}, {\it arXiv:2205.15204 [cs.PL]}.

\bibitem[Liu et~al., 2023b]{Liu+23RuleLangBench-ICLP}
{\sc Liu, Y.~A.}, {\sc Stoller, S.~D.}, {\sc Tong, Y.}, {\sc and} {\sc Tekle,
  K.~T.}
\newblock Benchmarking for integrating logic rules with everything else.
\newblock In {\em Proceedings of the 39th International Conference on Logic
  Programming (Technical Communications)} 2023b, pp. 12--26. Open Publishing
  Association.

\bibitem[Loo et~al., 2009]{loo09decl}
{\sc Loo, B.~T.}, {\sc Condie, T.}, {\sc Garofalakis, M.}, {\sc Gay, D.~E.},
  {\sc Hellerstein, J.~M.}, {\sc Maniatis, P.}, {\sc Ramakrishnan, R.}, {\sc
  Roscoe, T.}, {\sc and} {\sc Stoica, I.} 2009.
\newblock Declarative networking.
\newblock {\em Communications of the ACM}, {\it 52}, 11, 87--95.

\bibitem[Ma et~al., 2006]{ma2006towards}
{\sc Ma, L.}, {\sc Yang, Y.}, {\sc Qiu, Z.}, {\sc Xie, G.}, {\sc Pan, Y.}, {\sc
  and} {\sc Liu, S.}
\newblock Towards a complete {OWL} ontology benchmark.
\newblock In {\em European Semantic Web Conference} 2006, pp. 125--139.
  Springer.

\bibitem[Maier et~al., 2018]{maier18hist-wbook}
{\sc Maier, D.}, {\sc Tekle, K.~T.}, {\sc Kifer, M.}, {\sc and} {\sc Warren,
  D.~S.}
\newblock Datalog: {Concepts}, history and outlook.
\newblock In {\sc Kifer, M.} {\sc and} {\sc Liu, Y.~A.}, editors, {\em
  Declarative Logic Programming: Theory, Systems, and Applications} 2018,
  chapter~1, pp. 3--120. ACM and Morgan \& Claypool.

\bibitem[Marques-Silva and Sakallah, 1999]{marques1999grasp}
{\sc Marques-Silva, J.~P.} {\sc and} {\sc Sakallah, K.~A.} 1999.
\newblock Grasp: A search algorithm for propositional satisfiability.
\newblock {\em IEEE Transactions on Computers}, {\it 48}, 5, 506--521.

\bibitem[McAllester, 1999]{mcallester99}
{\sc McAllester, D.~A.}
\newblock On the complexity analysis of static analyses.
\newblock In {\em Proceedings of the 6th International Static Analysis
  Symposium} 1999, pp. 312--329. Springer.

\bibitem[Qureshi and Faber, 2024]{qureshi2024evaluating}
{\sc Qureshi, H.~M.} {\sc and} {\sc Faber, W.} 2024.
\newblock Evaluating {Datalog} tools for meta-reasoning over {OWL 2 QL}.
\newblock {\em Theory and Practice of Logic Programming}, {\it 24}, 2,
  368--393.

\bibitem[Rothamel and Liu, 2007]{RotLiu07Retrieval-PEPM}
{\sc Rothamel, T.} {\sc and} {\sc Liu, Y.~A.}
\newblock Efficient implementation of tuple pattern based retrieval.
\newblock In {\em Proceedings of the ACM SIGPLAN 2007 Workshop on Partial
  Evaluation and Program Manipulation} 2007, pp. 81--90.

\bibitem[Sagonas et~al., 1994]{SagSW94xsb}
{\sc Sagonas, K.}, {\sc Swift, T.}, {\sc and} {\sc Warren, D.~S.}
\newblock {XSB} as an efficient deductive database engine.
\newblock In {\em Proceedings of the 1994 ACM SIGMOD Intl.\ Conf.\ Management
  of Data} 1994, pp. 442--453.

\bibitem[Seo et~al., 2013]{seo2013socialite}
{\sc Seo, J.}, {\sc Guo, S.}, {\sc and} {\sc Lam, M.~S.}
\newblock Socialite: {Datalog} extensions for efficient social network
  analysis.
\newblock In {\em 2013 IEEE 29th International Conference on Data Engineering}
  2013, pp. 278--289.

\bibitem[Singh et~al., 2020]{singh2020owl2bench}
{\sc Singh, G.}, {\sc Bhatia, S.}, {\sc and} {\sc Mutharaju, R.}
\newblock {OWL2Bench}: a benchmark for {OWL 2} reasoners.
\newblock In {\em International semantic web conference} 2020, pp. 81--96.
  Springer.

\bibitem[Smaragdakis and Balatsouras, 2015]{smara15pointer}
{\sc Smaragdakis, Y.} {\sc and} {\sc Balatsouras, G.} 2015.
\newblock Pointer analysis.
\newblock {\em Foundations and Trends in Programming Languages}, {\it 2}, 1,
  1--69.

\bibitem[Souffle, 2026]{souffle26github}
Souffle 2026.
\newblock {The Souffl\'e Project}.
\newblock \url{https://github.com/souffle-lang}.
\newblock Accessed Feb.\ 27, 2026.

\bibitem[Sterling and Shapiro, 1994]{Sterling:Shapiro:94}
{\sc Sterling, L.} {\sc and} {\sc Shapiro, E.} 1994.
\newblock {\em The Art of Prolog}.
\newblock MIT Press, 2nd edition.

\bibitem[SWI-Prolog bench, 2026]{swi26bench}
SWI-Prolog bench 2026.
\newblock {SWI-Prolog benchmark suite}.
\newblock \url{https://github.com/SWI-Prolog/bench}.
\newblock Accessed Feb.\ 27, 2026.

\bibitem[Swift and Warren, 2012]{swift2012xsb}
{\sc Swift, T.} {\sc and} {\sc Warren, D.~S.} 2012.
\newblock {XSB}: {Extending Prolog} with tabled logic programming.
\newblock {\em Theory and Practice of Logic Programming}, {\it 12}, 1-2,
  157--187.

\bibitem[Swift et~al., 2022]{xsb22}
{\sc Swift, T.}, {\sc Warren, D.~S.}, {\sc Sagonas, K.}, {\sc Freire, J.}, {\sc
  Rao, P.}, {\sc Cui, B.}, {\sc Johnson, E.}, {\sc de~Castro, L.}, {\sc
  Marques, R.~F.}, {\sc Saha, D.}, {\sc Dawson, S.}, {\sc and} {\sc Kifer, M.}
  2022.
\newblock {\em The XSB System Version 5.0,x}.
\newblock \url{http://xsb.sourceforge.net}. Latest release October 29, 2017.

\bibitem[Tamaki and Sato, 1986]{TamSat86}
{\sc Tamaki, H.} {\sc and} {\sc Sato, T.}
\newblock {OLD} resolution with tabulation.
\newblock In {\em Proceedings of the 3rd International Conference on Logic
  Programming} 1986, volume 225 of {\em LNCS}, pp. 84--98. Springer.

\bibitem[Tekle et~al., 2008]{Tek+08RulePE-AMAST}
{\sc Tekle, K.~T.}, {\sc Hristova, K.}, {\sc and} {\sc Liu, Y.~A.}
\newblock Generating specialized rules and programs for demand-driven analysis.
\newblock In {\em Proceedings of the 12th International Conference on Algebraic
  Methodology and Software Technology} 2008, volume 5140 of {\em LNCS}, pp.
  346--361. Springer.

\bibitem[Tekle and Liu, 2010]{TekLiu10RuleQuery-PPDP}
{\sc Tekle, K.~T.} {\sc and} {\sc Liu, Y.~A.}
\newblock Precise complexity analysis for efficient {Datalog} queries.
\newblock In {\em Proceedings of the 12th International ACM SIGPLAN Symposium
  on Principles and Practice of Declarative Programming} 2010, pp. 35--44.

\bibitem[Tekle and Liu, 2011]{TekLiu11RuleQueryBeat-SIGMOD}
{\sc Tekle, K.~T.} {\sc and} {\sc Liu, Y.~A.}
\newblock More efficient {Datalog} queries: {S}ubsumptive tabling beats magic
  sets.
\newblock In {\em Proceedings of the 2011 ACM SIGMOD International Conference
  on Management of Data} 2011, pp. 661--672.

\bibitem[Vianu, 2021]{vianu2021datalog}
{\sc Vianu, V.}
\newblock Datalog unchained.
\newblock In {\em Proceedings of the 40th ACM SIGMOD-SIGACT-SIGAI Symposium on
  Principles of Database Systems} 2021, pp. 57--69.

\end{thebibliography}
\arxiv{}
\tlp{
\bibliographystyle{tlplike}
}
}
} %

}

\aaai{}

\sepexpe{
\appendix
\tlp{\clearpage}
\section{Proofs for optimal number of combinations in Table~3}
\label{sec-calc}

\newcommand{\mygtype}{\Vex{2}\noindent}


We give proofs for the formulas in Table~3 for optimal number of combinations for the recursive rule for all 3 recursion variants---%
left recursion (\co{path(X,Y), edge(Y,Z)}),
right recursion (\co{edge(X,Y), path(Y,Z)}),
and double recursion (\co{path(X,Y), path(Y,Z)})---%
and all graph types.
The proofs use the numbers of edges and paths, and of values of variables
(\co{X}, \co{Y}, and \co{Z}) holding the vertices.

\mygtype
\gname{Cmpl}{n}, with edges \co{\{(i,j):~i in 1..n, j in 1..n\}}

Given \co{edge} holds for all pairs of vertices.
By base case rule of \co{path}, \co{path} holds for all of them too.
Therefore, for all 3 recursion variants, all combinations of \m{n} vertices
for all three variables must be counted, yielding \m{n^3} combinations, as
shown in Table~3.

\mygtype
\gname{MaxAcyc}{n}, with edges \co{\{(i,j):~i in 1..n-1, j in i+1..n\}}

Given \co{edge} holds for all vertex pairs \m{(i,j)} where \m{i < j}.
By base case rule of \co{path}, \co{path} holds for all of them too.
Therefore, for all 3 recursion variants, 
for each vertex \m{i} from 1 to \m{n} for variable \co{Y},
there is a path or edge from each vertex from 1 to \m{i-1} for \co{X}
and a path or edge to each of vertex \m{i+1} to \m{n} for \co{Z}.
So the number of combinations is \m{\sum_{i = 1..n} (i-1) (n-i)}, 
yielding the closed form in Table~3.

\mygtype
\gname{Cyc}{n}, with edges \co{\{(i,i+1):~i in 1..n-1\} + \{(n,1)\}}

Given \co{edge} holds for all vertex pairs 
\m{(i,i+1)} for \m{i} from 1 to \m{n-1} and for pair \m{(n,1)}.
By transitivity, \co{path} holds for all pairs of vertices.
For left (resp.\ right) recursion, for each \co{path} pair \m{(i,j)},
there is only one edge going from \m{j} (resp.\ to \m{i}), and thus the 
number of combinations is the size of \co{path}, which is \m{n^2},
as shown in Table~3.
For double recursion, the analysis is as for \gname{Cmpl}{n}, yielding
\m{n^3} in Table~3.

\mygtype
\gname{CycExtra}{n,k}, with edges 
\co{\{(i,(i-1 + t*n/(k+1)) mod n + 1):~i in 1..n, t in 1..k\} + \gname{Cyc}{n}}

This is as \gname{Cyc}{n} but adds edges from each vertex to \m{k} vertices,
so there are \m{n+nk} edges total.
Still, \co{path} holds for all pairs of vertices.
For left recursion, for each \co{path} pair \m{(i,j)}, there are \m{k} more
edges going from \m{j}, and thus the number of combinations is \m{n^2+n^2 k},
as shown in Table~3.
For right recursion, for each of \m{n} + \m{nk} edges, there are paths to
\m{n} vertices, and thus the number of combinations is again \m{n^2+n^2 k},
as shown in Table~3.
For double recursion, the analysis is the same as for \gname{Cyc}{n,k}.

\mygtype
\gname{Path}{n}, with edges \co{\{(i,i+1):~i in 1..n-1\}}

Given \co{edge} holds for all vertex pairs \m{(i,i+1)} for \m{i=1..n}.
By transitivity, \co{path} holds for all pairs \m{(i,j)} where \m{i<j},
so the number of \co{path} pairs is \m{\sum_{i = 1..n} (n-i)}, which equals
\m{\frac{1}{2} (n - 1) (n - 2)}.
For left (resp.\ right) recursion, for each \co{path} pair \m{(i,j)}, 
there is one edge going from \m{j} (resp.\ to \m{i}), 
and thus the number of combinations is the number of \co{path} pairs,
i.e., \m{\frac{1}{2} (n - 1) (n - 2)}, as in Table~3.
For double recursion, the calculation is the same as for
\gname{MaxAcyc}{n}, yielding the closed form in Table~3.

\mygtype
\gname{PathDisj}{n,k}, with edges \co{\{(i,i+k):~i in 1..(n-1)*k\}}

This is the same as \gname{Path}{n}, but multiplied by \m{k} 
for each recursion variant.

\mygtype
\gname{Grid}{n}, with edges
\begin{tabular}[t]{@{\tlp{\Hex{-42}}}l}
\co{\{(j, j+1):~i in 1..n, j in (i-1)*n+1..i*n-1\} +}\\
\co{\{(j, j+n):~i in 1..n-1, j in (i-1)*n+1..i*n\}}  
\end{tabular}

For left recursion, 
for each vertex in position \m{(i,j)} of the grid not in the last row or column,
there are paths from \m{i\times j-1} vertices before and edges to 2
vertices after;
for each vertex in the last column on row \m{i} but not last row,
there are paths from \m{i\times n -1} vertices and an edge to 1 vertex;
for each vertex in the last row but on column j not last column,
there are paths from \m{n\times j -1} vertices and an edge to 1 vertex.
Thus, the number of combinations is 
\m{\sum_{i = 1..n-1, j = 1..n-1}  (i\times j -1)\times 2 +
   \sum_{i = 1..n-1} (i\times n -1) +
   \sum_{j = 1..n-1} (n\times j -1)},
yielding the closed form in Table~3.

For right recursion, the counting is similar, except to 
consider each vertex not in the first row or column,
then each vertex in first column but not first row,
and each vertex in first row but not first column,
edges from 2 or 1 vertex before, and paths to vertices after.
Thus, the number of combinations is
\m{\sum_{i = 2..n, j = 2..n}  2\times ((n-i+1)\times(n-j+1) -1) +
   \sum_{i = 2..n} ((n-i+1)\times n -1) +
   \sum_{j = 1..n} (n\times (n-j+1) -1) },
yielding the closed form in Table~3.

For double recursion, 
for each vertex in position \m{i=1..n} and \m{j=1..n} in the grid,
there are paths from \m{i\times j -1} vertices before,
and paths to \m{(n-i+1)\times(n-j+1) -1} vertices after.
Thus the number of combinations is 
\m{\sum_{i = 1..n, j = 1..n} (i\times j - 1)\times ((n-i+1)\times(n-j+1) -1)},
yielding the closed form in Table~3.

\mygtype
\gname{BinTree}{h}, with edges
\co{\{(i,2i):~i in 1..2\m{^{\tt{h-1}}}\} +
\{(i,2i+1):~i in 1..2\m{^{\tt{h-1}}}\}}
and\\ 
\gname{BinTreeRev}{h}, with edges
\co{\{(2i,i):~i in 1..2\m{^{\tt{h-1}}}\} +
\{(2i+1,i):~i in 1..2\m{^{\tt{h-1}}}\}}

Left and right recursions are as proved in Section~4.1.
Double recursion for \gname{BinTree}{h} (resp. \gname{BinTreeRev}{h})
considers each level \m{i=1..h} from root to leaves,
with \m{2^{i-1}} vertices at level \m{i}, each having paths from (resp.\ to)
\m{i-1} vertices, and to (resp.\ from) \m{2^{h-i+1} -2} vertices. 
Thus the number of combinations is
\m{\sum_{i = 1..h} 2^{i-1}\times (i-1)\times (2^{h-i+1} -2)},
yielding the closed form in Table~3.

\mygtype
\gname{X}{n,k}, with edges
\co{\{(i,n+1):~i in 1..n\} + \{(n+1,n+1+j):~j in 1..k\}}

All 3 recursion variants just join the two sets of edges,
and thus the number of combinations is \m{nk}, as shown in Table~3.

\mygtype
\gname{Y}{n,k}, with edges
\co{\{(i,n+1):~i in 1..n\} + \{(i,i+1):~i in n+1..n+k-1\}}

For left recursion, 
each vertex \m{i} from 1 to \m{k-1} in the stem of Y has paths 
from \m{n+i-1} vertices.
Thus the number of combinations is 
\m{\sum_{i = 1..k-1} (n+i-1)}, yielding the closed form in Table~3.

For right recursion,
vertex \m{n+1} has edges from \m{n} vertices and paths to \m{k-1} vertices;
each vertex \m{i} from 2 to \m{k-1} in the stem of Y has edge from one vertex
and paths to \m{k-i} vertices.
Thus the number of combinations is 
\m{n\times (k-1) + \sum_{i = 2..k} (k-i)}, yielding the closed form in Table~3.

For double recursion,
each vertex \m{i} as in left recursion has also paths from \m{n+i-1} vertices,
but has paths to \m{k-i} vertices.
Thus, the number of combinations is 
\m{\sum_{i = 1..k-1} (n+i-1)\times (k-i)}, 
yielding the closed form in Table~3.

\mygtype
\gname{W}{n,k}, with edges
\co{\{(i,n+1 + (i+j-1) mod n):~i in 1..n, j in 1..k\}}

This graph has only paths of length 1.
All 3 recursion variants have the number of combinations being 0,
as shown in Table~3.

\aaai{}
\mysec{Performance measurements and analysis}
\label{sec-app}

\newcommand{\floor}[1]{\left\lfloor #1 \right\rfloor}

We designed and performed detailed experiments with all three rule systems
XSB, clingo, and \souf, all three rule variants, and all 12 input graph
types.\footnote{Our benchmarking experiments
  are available at~\url{https://github.com/Sirneij/trans-bench}.}  We
describe these experiments, analyze the measured running times, and compare
with the analyzed optimal complexities.
\begin{description}

\item For XSB, with the general programming power of Prolog, we wrote the
  scripting and timing code directly in XSB.
  We obtain separate times for (1) \co{LoadRules}---load 
  rules (including \co{table} directive), using \co{consult},
  (2) \co{ReadData}---read data from a file of facts, using \co{load\_dync}, %
  (3) \co{Query}---query all \co{path} facts, using \co{fail} to repeat without
  writing, and
  (4) \co{WriteRes}---write results to file, using \co{write}. 

\item For clingo, with its interface for using it from Python, we
  wrote the scripting and timing code in Python.
  We obtain separate times for (1) \co{LoadRules}---load rules (including
  \co{show} directive), using \co{load}, (2) \co{ReadData}---read data from
  a file of facts, using \co{load}, (3) \co{Ground}---ground rules, using
  \co{ground}, and (4) \co{Solve}---search for all \co{path} facts, using
  \co{solve}, and (5) \co{WriteRes}---write results to file, using
  \co{write}.

\item For \souf, with its interface for using it from C++, we wrote the
  scripting and timing code in C++.
  We obtain separate times for (1) \co{RulesToC++}---compile rules (with
  type and input-output declarations) to C++, using \co{souffle} command
  line,
  (2) \co{C++ToExec}---compile C++ to executable, using 
  \co{g++} command line,
  (3) \co{LoadRules}---load compiled executable for rules, 
  using \co{newInstance},
  (4) \co{ReadData}---read data from a directory of files of facts, one for
  each given predicate, using \co{loadAll},
  (5) \co{Query}---compute all path facts, using \co{run}, and
  (6) \co{WriteRes}---write results to file in CSV format, using
  \co{printAll}.



\end{description}
For each input graph type, we generate graphs, exactly as defined in
Table~\ref{tab-graphs}, with increasing values for \co{n} and \co{h}, and
with both constant values and increasing values for \co{k}.

For each experiment, we report average CPU time over 5 runs.
The variations across runs are negligible.
All measurements were taken on 
\oldtlp{%
an Apple M2 chip, 2022, with a 
3.50 GHz 8-core CPU and 8 GB RAM,
running MacOS Sonoma 14.4.1, Python 3.12, XSB 5.0.0, clingo 5.7.1, \souf
2.4.1-26-gc7ce22981, Python 3.12.0, and Apple clang 15.0.0 (clang-1500.3.9.4).
}%
an Apple M3 Pro chip, 2024, with a 4.05 GHz 11-Core CPU and 18 GB unified
memory, 512 GB SSD, running MacOS Sonoma 14.5, 
XSB 5.0.0, clingo 5.7.1, and \souf 2.4.1-50-ge6cc66820.

\mysubsec{Running times, confirmations, and exceptions}

We discuss running times first for graph types with a single parameter,
\co{n} or \co{h}, and then for graph types with two parameters, \co{n} and
\co{k}, but we compare across all graph types, and for all recursion
variants and all systems.

\mypar{Graph types with a single parameter}
We first examine the running times for five drastically different graph
types with parameter \co{n} for number of vertices: \gname{Cmpl}{n},
\gname{Cycle}{n}, \gname{MaxAcyc}{n}, \gname{Path}{n}, and \gname{Grid}{n},
including densest (\gname{Cmpl}{} graphs) and sparsest connected
(\co{Path}{} graphs).
We explain how the calculated precise complexities in Table~\ref{tab-calc}
and the different optimized inference efficiencies in
Section~\ref{sec-difftime} are confirmed by the sameness and differences as
well as trends and constants in actual inference times---\co{Query} for XSB
and \souf and \co{Ground} and \co{Solve} for clingo.  The input and output
times are also explained by data and result sizes, but they are relatively
small here, so we focus on inference times.

\plotfig{Cmpl}{1000000}{1000}
\plotfig{Cyc}{1000}{1000}

Figs.~\ref{fig-Cmpl} to~\ref{fig-Grid} show the running time details for
\gname{Cmpl}{n}, \gname{Cycle}{n}, \gname{MaxAcyc}{n}, \gname{Path}{n}, and
\gname{Grid}{n}, for the number of vertices up to 1000 and the number of
edges varying the mostly widely possible, from 1000000 (for
\gname{Cmpl}{1000}) to 999 (for \gname{Path}{1000}) for a graph to stay
connected.  First, we observe the following that hold across all Figures,
not just the five figures here---all consistent with the analysis in
Section~\ref{sec-difftime}---except for an XSB performance bug discovered
on the simplest case (\gname{W}{} graphs, left recursion), described later:
\begin{itemize}
\item \souf has an extra compilation time (about 1.5 seconds, with tiny
  \co{RulesToC++}, in yellow, and virtually all \co{C++ToExec}, in orange).
  Of course, this time needs to be incurred only once for each program, not
  each run of the program.

\item clingo's grounding time (\co{Ground}, in slightly lighter red)
  dominates its solving time (\co{Solve}) which is tiny, because optimized
  grounding already inferred all resulting facts.

\item XSB is the fastest in all runs, confirming the overhead of tabling is
  minimal compared with the cost of grounding in clingo and semi-naive
  evaluation in \souf.
\end{itemize}
We also confirmed the trends of being superlinear in \co{n} for these graph
types, as well as, for all other graph types, the expected trends of being
linear or superlinear in \co{n} and \co{k} and of an explosive increase
with respect to \co{h}, as analyzed in Table~\ref{tab-calc}, except for the
case where the XSB performance bug was discovered.

\arxiv{
\plotfig{MaxAcyc}{499500}{1000}
\plotfig{Path}{999}{1000}
}

For more interesting detailed comparison among different recursions with
different inference methods, we first examine Fig.~\ref{fig-Cmpl}, for
\gname{Cmpl}{n}. It is the most expensive case, with 1,000,000,000
combinations for 1000 vertices and 1,000,000 edges.
Table~\ref{tab-calc} says that all three recursion types have the same
complexity, but Fig.~\ref{fig-Cmpl} shows it is only partially the case,
and the difference confirms the analysis in Section~\ref{sec-difftime}:
\begin{itemize}

\item For both XSB and clingo, right and double recursions have about the
  same running times, but slower than left recursion (by about 40\% for XSB
  and 220\% for clingo).
  %
  The exceptions are actually consistent with the analysis in
  Section~\ref{sec-difftime}: left recursion is faster because both
  consider hypotheses of rules from left to right, and clingo is slower
  because it uses ground rules and semi-naive evaluation.

\item For \souf, left and right recursions have about the same running
  times, consistent with the analysis in Section~\ref{sec-difftime}, but double
  recursion is twice as slow.
  Indeed, with semi-naive, 
  considering all new \co{path} facts in each iteration for double
  recursion
  has to essentially process each \co{path} fact twice, compared with only
  once for left or right recursion.

\end{itemize}
\tlp{
\plotfig{MaxAcyc}{499500}{1000}
\plotfig{Path}{999}{1000}
}
Examining Figs.~\ref{fig-Cyc} to~\ref{fig-Path} for other graph types
shows and confirms additional details.
\begin{itemize}
\item For \gname{Cycle}{n}, Table~\ref{tab-calc} says that double
  recursion has the same complexity as \gname{Cmpl}{n},
  whereas left and right recursions are a linear factor faster.
  Fig.~\ref{fig-Cyc} indeed supports those for XSB and \souf, compared
  with Fig.~\ref{fig-Cmpl} for \gname{Cmpl}{n}.
  For clingo, it is faster in Fig.~\ref{fig-Cyc}(c) than in
  Fig.~\ref{fig-Cmpl}(c),
  an effect of grounding over sparser input.
  \oldtlp{The only other main difference is that Fig.~\ref{fig-Cyc} has tiny
  \co{ReadData} time compared with Fig.~\ref{fig-Cmpl}.}

\item For \gname{MaxAcyc}{n}, Table~\ref{tab-calc} says that all three
  recursion types have the same complexity, about
  1/6 of that of \gname{Cmpl}{n}.
  %
  Fig.~\ref{fig-MaxAcyc} confirms that for XSB and \souf, including
  the constant 1/6.
  For example, \souf on double recursion takes about \oldtlp{4}54 seconds
  (\oldtlp{5.5-1.5}55.5-1.5) at \oldtlp{400}1000 vertices here, and about
  \oldtlp{24}322 seconds (\oldtlp{25-1.5}323.5-1.5) in
  Fig.~\ref{fig-Cmpl}.
  For clingo, slightly sparser graphs here have slightly less running times
  than in Fig.~\ref{fig-Cmpl}.

\item For \gname{Path}{n},
  the relationship of Fig.~\ref{fig-Path} to Fig.~\ref{fig-MaxAcyc} 
  is exactly similar as
  the relationship of Fig.~\ref{fig-Cyc} to Fig.~\ref{fig-Cmpl}.
\end{itemize}

\plotfig{Grid}{1860}{\m{\floor{\sqrt{1000}}^2} (= 961,
closest from below to form a grid, and calculated this way for all x-axis
labels)}
\plotfig{BinTree}{510}{\m{2^{\floor{\log_2 1000}}-1} (= 511, closest from
  below to 1000 to form a binary tree, and calculated this way for all
  x-axis labels)}
\plotfig{BinTreeRev}{510}{\m{2^{\floor{\log_2 1000}}-1}}

Similar analysis as above applies to \gname{Grid}{n}, where \co{n} is (the
closest from below to) the square root of the number of vertices.  The
running times are shown in Fig.~\ref{fig-Grid}. The main difference from
first four graph types is that the inference times are noticeably smaller,
and that other times besides inference times are more visible.  This is due
to much shorter paths to follow (shorter than the square root of \co{n}).

Similar analysis applies also to \gname{BinTree}{h} and
\gname{BinTreeRev}{h} with parameter \co{h} for level of nodes, except that
paths are even drastically shorter (logarithm of the number of nodes).
The running times, shown in Figs.~\ref{fig-BinTree}
and~\ref{fig-BinTreeRev}, are so small that only the compilation time shows
distinctively.  We also experimented with trees with drastically more nodes
and observed that the running times increase explosively with respect to
\co{h}, and that both tree types have about the same running times, all
consistent with analyzed.

\mypar{Graph types with two parameters}

The remaining five graph types---\gname{CycExtra}{n,k}, \gname{PathDisj}{n,k},
\gname{X}{n,k},
\gname{Y}{n,k}, and
\gname{W}{n,k}---have two parameters \co{n} and \co{k}.  We
discuss results with two sets of experiments, both with increasing \co{n}
up to 1000:
(1) with constant \co{k} = 10 for all five graph types, and 
(2) with increasing \co{k} = \co{n} for the last three
graph types.
The latter helps greatly in showing the different impacts on performance by
the last three graph types.  It also led us discover an XSB performance bug
in the simplest case of left recursion on \gname{W}{} graphs.

\plotfig{CycExtra}{11000}{1000}
\plotfig{PathDisj}{9990}{10000}
Figs.~\ref{fig-CycExtra} to~\ref{fig-w} show the running times for all
five kinds of graphs for constant \co{k} = 0.  They confirm the analysis as
before and additionally the constant factor differences except for what we
believe is a good cache locality.
\begin{itemize}
\item Comparing Fig.~\ref{fig-CycExtra} for \gname{CycExtra}{n,10} with
  Fig.~\ref{fig-Cyc} for \gname{Cyc}{n}, we can see they are similar for
  double recursion, consistent with the analysis in
  Section~\ref{sec-analyze}.  Running times for left and right recursions
  are too small in these figures, but they generally are about 4-5 times
  for XSB and \souf and about 3 times for clingo
  (e.g., from 0.10 to 0.45 seconds for XSB,
  from 0.66 to 1.85 seconds for clingo, and
  from 1.28 to 6.17 seconds for \souf, 
  for left recursion on 1000 vertices).

  These constant factors are smaller than the 10-times-larger analyzed in
  Section~\ref{sec-analyze}.  We believe this is due to the good cache
  locality of 10 added edges starting from a same vertex in
  \gname{CycExtra}{n,10}, contrasting each edge starting from a different
  vertex in \gname{Cyc}{n}.

\item Comparing Fig.~\ref{fig-PathDisj} for \gname{PathDisj}{n,10} with
  Fig.~\ref{fig-Path} for \gname{Path}{n}, we can clearly see that the
  running times are about 10 times as large for double recursion, again
  with analysis in Section~\ref{sec-analyze}.  For left and right
  recursions, this is true too even though the running times are again too
  small to see in the figures (e.g., from 0.05 to 0.59 seconds for XSB,
  from 0.33 to 3.78 seconds for clingo, and 
  from 0.60 to 6.31 seconds for \souf). 
  Indeed, each \gname{PathDisj}{n,10} graph is exactly the same as 10
  \gname{Path}{n} graphs.

\item Figures
  for the last three graph types show that the running times for \co{k} =
  10 are all so small that the compilation time stands out again.  But
  those small running times do show different coloring, prompting the next
  set of experiments.
\end{itemize}

\plotfig{X}{1010}{1011}
\plotfig{Y}{1009}{1010}
\plotfig{w}{10000}{2000}

Figs.~\ref{fig-X} to~\ref{fig-W} show the running times for the last
three graph types for \co{k} = \co{n}, i.e., \co{k} increases as \co{n}
increases.  Fig.~\ref{fig-Y} for \gname{Y}{n,n} can be analyzed similarly
as for
\gname{Path}{n},
but the most interesting cases are Figs.~\ref{fig-X} and~\ref{fig-W} for
\gname{X}{n,k} and \gname{W}{n,k}, respectively.

\begin{itemize}

\item Fig.~\ref{fig-X} for \gname{X}{n,n} shows two new aspects clearly:
  (1) \co{WriteRes} times (in blue) are totally obvious, and they are much
  larger for XSB and clingo than for \souf; and (2) \co{Solve} times (in
  darker red) in clingo are obvious on top of \co{Ground} times (in lighter
  red).

  Fig.~\ref{fig-X} also shows an exception from before: inference time
  for left recursion got worse, relative to right recursion, for all of
  XSB, clingo, and \souf. XSB is about 150\% slower, instead of faster;
  clingo is about 1\% faster, instead of much faster;
  \souf is about 50\% slower, instead of the same.
  Note though that these variations are relatively very small compared with
  the times for writing results by XSB and clingo and compilation by \souf.

\ploteqfig{X}{2000}{2001}
\ploteqfig{Y}{1999}{2000}
\ploteqfig{W}{1000000}{2000}

  The exception is due to the specifics of \gname{X}{} graphs having only
  paths of lengths 1 and 2, where faster table lookups for \co{path} pairs
  in left recursion is overwhelmed by the sheer number of them that fail
  anyway.
  In particular, \gname{X}{n,n} has \co{\#edge} = \co{2n} and \co{\#path} =
  \co{2n+n\m{^2}}.  Since XSB and clingo consider hypotheses from left to
  right, left recursion looks up \co{2n+n\m{^2}} \co{path} pairs first, and
  only \co{n} of them succeed, whereas right recursion looks up \co{2n}
  \co{edge} pairs first, and \co{n} of them succeed.  We believe this
  difference in the number of lookups is also the case for \souf.  Optimal
  computation with minimum increment can achieve the same efficiency for
  all recursion types.

\item Fig.~\ref{fig-W} for \gname{W}{n,n} shows one more new aspect:
  \co{ReadData} times (in green) are totally obvious, and they are about
  twice as high for XSB and clingo than for \souf.

  The biggest surprise is the discovery of an XSB performance bug: The
  running times of XSB on left recursion grow drastically higher than
  clingo and \souf and than analyzed.  This is actually the simplest case
  among all graph types we consider, where \co{path} is exactly \co{edge}
  with no other paths at all.

  This performance bug was confirmed by the XSB team.  They then analyzed
  and determined that it is due to hash collisions in this case.  They also
  showed that additional, more sophisticated indexing directives in XSB
  could be added to the program to bypass this bug.

\end{itemize}

%
%



}

\end{document}